\documentclass[pra,a4paper,twocolumn,floatfix,superscriptaddress]{revtex4}
\usepackage{pstricks,graphicx,amsmath,bbm,mathrsfs,amssymb,times, verbatim}
\newcommand{\ket}[1]{\vert #1 \rangle}

\def\var{{\rm var}}
\def\omegaA{\omega_{\rm A}}

\def\HA{{\cal H}_{\rm A}}
\def\Hint{{\cal H}_{\rm int}}
\def\sigmaX{\sigma_{x}}
\def\sigmaY{\sigma_{y}}
\def\sigmaZ{\sigma_{z}}
\def\varepsilonF{\vec{\varepsilon}_{\rm F}}

\def\Fri{F_{\omegaA}^{\rm (Rbi)}}

\def\Fry{F_{\omegaA}^{\rm (Rmy)}}

\def\QF{H}

\def\pe{p_{\omegaA}^{(e)}}

\def\pg{p_{\omegaA}^{(g)}}
\def\pk{p_{\omegaA}^{(k)}}
\def\Pe{P_{\omegaA}^{(e)}}

\def\Pk{P_{\omegaA}^{(k)}}

\def\C{{\rm C}}
\def\NC{{\rm NC}}
\def\CPT{{\rm CPT}}
\def\gammaSte{{\widetilde{\gamma}}}
\def\GammaSte{{\widetilde{\Gamma}_p}}
\def\DeltaSte{{\widetilde{\Delta}}}
\def\FCPT{F_{\omegaA}^{\rm (CPT)}}
\def\FCPTen{F_{\omegaA|1,2,3}^{\rm (CPT)}}
\def\HCPT{H_{\omegaA}^{\rm (CPT)}}
\def\FCPTlev{F_{\omegaA| l}^{\rm (CPT)}}

\def\hvarrho{\hat{\varrho}}
\def\hL{{\cal L}}
\begin{document}
\title{The ultimate bounds to precision of atomic clock frequency measurement techniques}
\author{Stefano Olivares}
\email{stefano.olivares@fisica.unimi.it}
\affiliation{Dipartimento di Fisica ``Aldo Pontremoli'', Universit\`a degli Studi
di Milano and INFN Sezione di Milano, I-20133 Milano, Italia}
\author{Salvatore Micalizio}
\email{s.micalizio@inrim.it}
\affiliation{Istituto Nazionale di Ricerca Metrologica, INRIM, Quantum Metrology and Nanotechnologies Division, Strada delle Cacce 91, Torino, 10135, Italy}
\author{Matteo G.~A.~Paris}
\email{matteo.paris@fisica.unimi.it}
\affiliation{Dipartimento di Fisica ``Aldo Pontremoli'', Universit\`a degli Studi
di Milano and INFN Sezione di Milano, I-20133 Milano, Italia}
\date{\today}
\begin{abstract}
We investigate the ultimate quantum limits to the achievable uncertainty in the estimation of the transition frequency between two atomic levels. We focus on Rabi, Ramsey, and coherent population trapping (CPT) techniques, which are widely employed in experiments. We prove that in the Rabi and Ramsey schemes measuring the atomic population allows one to reach the minimum uncertainty, but, for the CPT setup, a measurement involving the coherences between the levels results in a further improvement of the estimation. As a figure of merit, we consider the Fisher information of the population measurement and compare its value to the quantum Fisher information, corresponding to the maximum precision, optimized over all the possible feasible measurements. 
\end{abstract}
\maketitle
\section{Introduction}\label{s:introduction}
Estimating the frequency of an atomic transition is at the basis of a large variety of physical measurements, including frequency metrology \cite{riehle2004, vanier_audoin}, high resolution spectroscopy \cite{jiang_2011} and possible time variation of the fundamental constants \cite{safronova}. Several techniques have been developed to observe atomic transitions, particularly those characterized by a high signal-to-noise ratio and only partially influenced by broadening mechanisms. Moreover, to accurately determine the frequency of a transition with high precision, it must remain observable for an extended period, in accordance with the Heisenberg uncertainty principle. In this context, laser cooling techniques \cite{metcalf} have enabled the measurement of atomic clock frequencies with unprecedented accuracy, first using Cs/Rb fountains \cite{wynands_2005} and later with optical clocks \cite{ludlow2015}. 
\par
The optimization of an atomic frequency measurements can be tackled at a more fundamental level by means of the estimation theory. Among the available unbiased estimators, the one with the smallest variance offers the highest confidence in the accuracy of the estimate. In turn, a non-trivial lower bound on the variance may be defined by the Cramér-Rao inequality.
\par
Specifically, to estimate the (angular) frequency $\omega_A$ of a two-level atomic transition we start from a sample of atoms prepared in one of the two levels. Then, we let them interact with an oscillating electric field and a signal proportional to the transition probability $p_{\omega_A}(\omega)$ is maximized (or minimized) by changing the frequency $\omega$ of the applied field. This process leads to an estimate of $\omega_A$ through a measurement of $p_{\omega_A}(\omega)$. The framework of estimation theory imposes a bound on the uncertainty of the estimation of $\omegaA$, that is its variance, which  in this case satisfies the following inequality:
\begin{equation}
\var[\omegaA] \ge \frac{1}{F (\omegaA)} ,
\end{equation}
where $F (\omegaA)$ is the Fisher information \cite{helstrom}. Given that population measurement has only two possible outcomes, we have
\begin{align}
\ F (\omega)
=  \frac{\big[ \partial_{\omegaA} p_{\omega_A}(\omega)\big]^2}{p_{\omegaA}(\omega) \big[1-p_{\omegaA}(\omega) \big]}\,.
\label{Fisher}
\end{align}
Since the transition probability depends on the specific technique adopted to excite and detect the atomic transition, it is then of interest to investigate how the Fisher information expressed by Eq. (\ref{Fisher}) changes accordingly. 
\par
In this paper, we consider three techniques commonly used in atomic clock operation: i) the Rabi method, where the atoms interact with a single-mode electromagnetic field pulse; ii) the Ramsey scheme, where the interaction is split into two Rabi-like interactions separated by a much longer non-interacting region; iii) the coherent population trapping (CPT) phenomenon where the atoms interact with two phase coherent fields coupling the two clock transition levels to a common excited state. For each technique, we will calculate the Fisher information, and the corresponding quantum Fisher information \cite{helstrom,braun:QF,paris:QEQT}, i.e., the Fisher information achievable with the same physical probe and optimizing over all possible measurement allowed by quantum mechanics. 

In general, if $\hvarrho = \hvarrho({\lambda})$ is the density operator describing the state of the system under investigation
and $\lambda$ is the parameter we want to estimate, the minimum uncertainty is bounded by:
\begin{equation}
\var_{\rm min}[\lambda] \geq \frac{1}{\QF (\lambda)} ,
\end{equation}
where the $\QF (\lambda)$ is quantum Fisher information \cite{braun:QF}:
\begin{equation}
\QF (\lambda) = {\rm Tr} \Big[\hvarrho\,\hL_{\lambda}^2\Big],
\end{equation}
and we introduced $\hL_{\lambda}$, the symmetric logarithmic derivative (SLD), such that:
\begin{equation}
\hL_{\lambda}\,\hvarrho + \hvarrho\,\hL_{\lambda} = 2\, \partial_{\lambda}\hvarrho.
\end{equation}
If we write $\hvarrho = \sum_{n} r_n\, | \psi_n \rangle \langle \psi_n |$,
$\{ | \psi_n \rangle \}$ being the eigenbasis of $\hvarrho$ and $\{ r_n \}$
the corresponding eigenvalues, we have:
\begin{equation}\label{QFI:def}
\QF(\lambda) = 2\,\sum_{n,m}
\frac{\big| \langle  \psi_n  |  (\partial_\lambda \hvarrho) |  \psi_m  \rangle \big|^2}{r_n + r_m}\,,
\end{equation}
where the sum includes only the terms with $r_n + r_m \ne 0$.
\par
In the following, we show that estimation strategy based on the considered Rabi, Ramsey, and CPT techniques allow one to achieve the ultimate limits to precision imposed by quantum mechanics, since, at resonance, the corresponding Fisher and the quantum Fisher information coincide. Our approach based on estimation and quantum estimation theory leads us to prove that the Ramsey method beats both the Rabi and the CPT performance, as also suggested by the results of current experiments. Interestingly, when a measurement involving the coherences between the levels is considered, we found that for the CPT the Fisher and the quantum Fisher information are only slightly different for any value of the detuning $\Delta \omega = \omegaA - \omega$ and coincide at resonance; however, for the Rabi and Ramsey methods, the quantum Fisher information is sensibly higher than the classical one for nonzero values of $\Delta \omega$, thus fostering new investigation for different detection schemes allowing one to reach that limit.
\par
The paper is organized as follows. Sections~\ref{s:rabi} and \ref{s:ramsey} introduce the Rabi and Ramsey schemes, respectively, for the estimation of the atomic transition frequency together with the corresponding Fisher information, whereas the results concerning the quantum Fisher information are discussed in Section~\ref{s:QFI}.
The CPT technique is investigated in Section~\ref{s:CPT}, where we also calculate the related Fisher information and its quantum counterpart. We close the paper drawing some concluding remarks in Section~\ref{s:conclusion}.

\section{The Rabi method}\label{s:rabi}
Initially conceived for measuring nuclear magnetic moments \cite{rabi1938}, the Rabi resonance method soon became a relevant tool to investigate atomic spectra with high resolution. In this technique, the atoms interact with a single laser/microwave pulse, depending on the frequency of the transition considered. Observing the atoms' absorption as a function of the local oscillator frequency results in a discriminant curve that carries the information about the atomic resonance frequency. Rabi excitation is nowadays very often used in optical frequency standards
\cite{ludlow2015} because of its experimental simplicity compared to the Ramsey approach, as we will see later.

\begin{figure}[h!]
\includegraphics[width=0.9\columnwidth]{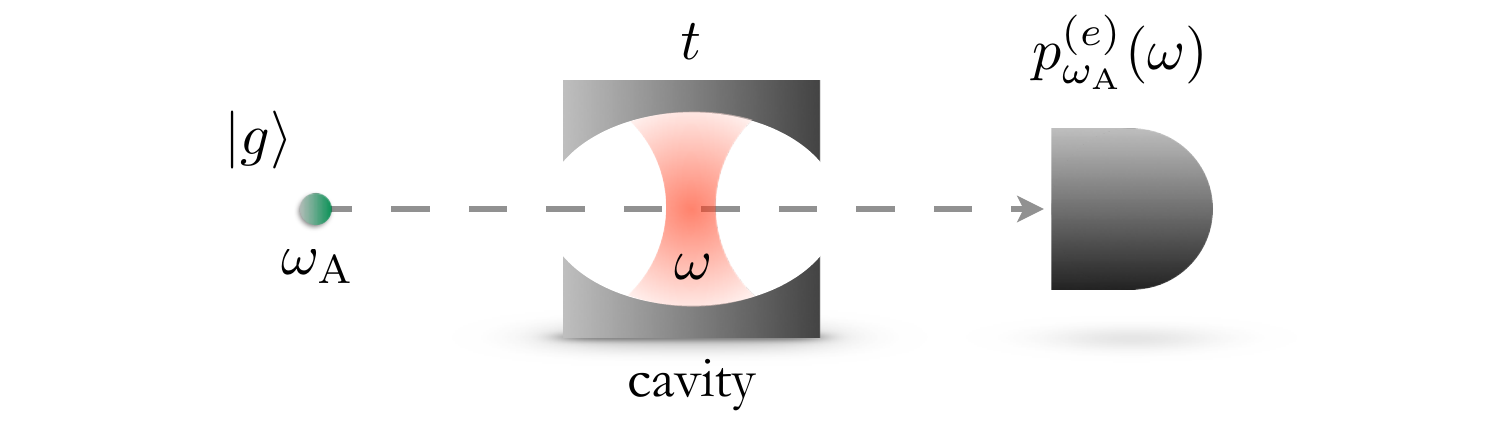}
\caption{Scheme of the Rabi method.}\label{f:rabi}
\end{figure}

\begin{figure}[h!]
\includegraphics[width=0.7\columnwidth]{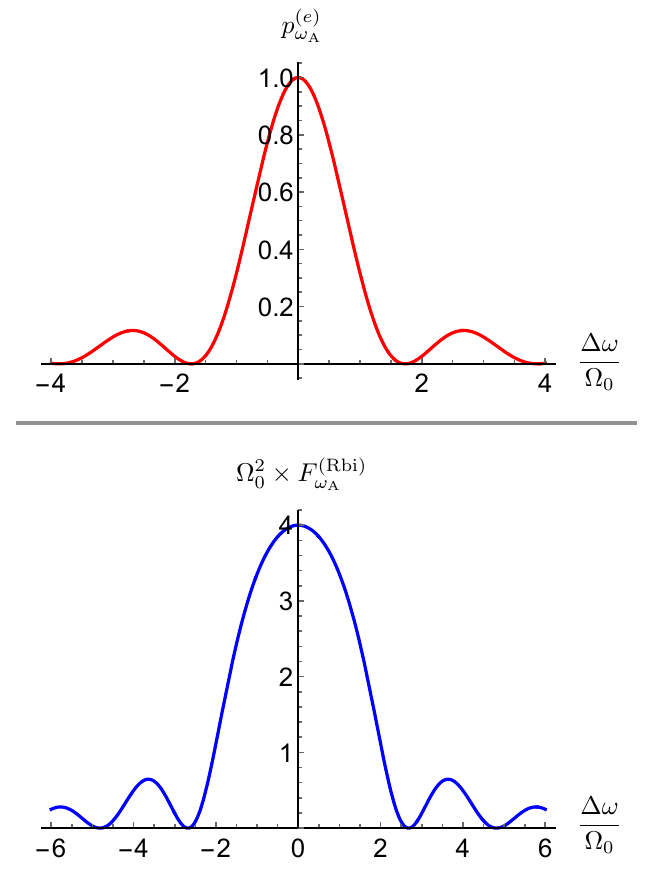}
\vspace{-0.2cm}
\caption{(Top) Plot of the Rabi probability $\pe$ to find the atom in the excited state after the interaction
with the cavity as a function of $\Delta\omega/\Omega_0$.
The atom is initially in the ground state and the interaction time corresponds to
a $\pi$-pulse.
(Bottom) Plot of the Fisher information $\Fri$ associated with the Rabi method as a
function of $\Delta\omega/\Omega_0$. The maximum is achieved at$\omega = \omegaA$. }\label{f:rabi:pe}
\end{figure}

The scheme of the Rabi method is sketched in Fig.~\ref{f:rabi}. A two-level atom, initially prepared in its ground state $\ket{g}$, interacts with an oscillating electric field inside a cavity for a time $t$. At resonance, i.e., when $\omega$ is equal to the atomic transition frequency $\omegaA$, if the interaction time is set to $t=\pi/\Omega_0$ (that is, in the presence of a $\pi$-pulse), $\Omega_0$ being the Rabi frequency, the atom is finally found in the excited state $\ket{e}$ with certainty. However, in the presence of a detuning $\Delta \omega = \omegaA - \omega$, the $\pi$-pulse brings the atom to its excited state with probability (see Appendix~\ref{app:Rabi} for details of the calculations):
\begin{equation}\label{eq:Rabi:pe}
\pe(\omega) = \frac{1}{\left[\Theta_{\Omega_0}(\Delta \omega)\right]^2}
\sin^2 \left[
\frac{\pi}{2}\, \Theta_{\Omega_0}(\Delta \omega)
\right]\,,
\end{equation}
with
\begin{equation}\label{Theta}
\Theta_{\Omega_0}(\Delta \omega) = 
\sqrt{1 + \left(\frac{\Delta \omega}{\Omega_0}\right)^2 }\,,
\end{equation}
which is plotted in the top panel of  Fig.~\ref{f:rabi:pe} as a function of $\Delta\omega/\Omega_0$.

It is clear that the maximum $\pe$ is attained at $\Delta\omega = 0$
and, thus, we can use $\pe$ to estimate the atomic transition frequency.
In this case, the uncertainty of the estimation of $\omegaA$, that is its variance,
satisfies the following inequality:
\begin{equation}
\var[\omegaA] \ge \frac{1}{\Fri (\omegaA)} ,
\end{equation}
where the $\Fri (\omegaA)$ is the Fisher information \cite{helstrom}, namely:
\begin{align}
\Fri (\omega) &= \sum_{k=g,e} \pk(\omega) \left[ \partial_{\omegaA} \ln \pk(\omega)\right]^2,\\
&=  \frac{\left[ \partial_{\omegaA} \pe(\omega)\right]^2}{\pe(\omega) \left[1-\pe(\omega) \right]}\,.
\label{F:rabi}
\end{align}
$\pg(\omega) = 1 - \pe(\omega)$ being the probability to detect the ground state of the atom after
the interaction with the cavity. The analytical expression of $\Fri (\omega)$ is quite clumsy and it is not reported here explicitly. We illustrate its behavior in the bottom panel of Fig.~\ref{f:rabi:pe} as a function of
$\Delta\omega/\Omega_0$. If $\omega \approx \omega_{\rm A}$, the following expansion holds:
\begin{equation}\label{Fri:expansion}
\Fri (\omega) \approx \frac{1}{\Omega_{0}^2} \left[
4 - \left( 8- \frac{3}{4}\, \pi^2 \right) ( \omega - \omega_{\rm A} )^2
\right]\,.
\end{equation}
\section{The Ramsey method}\label{s:ramsey}

The atomic resonance can be also probed by the Ramsey method of separated fields where the atoms interact with two phase coherent laser/microwave pulses separated by a dark interval \cite{ramsey1990}. Compared to the Rabi method, the Ramsey excitation provides a narrower Fourier-limited linewidth for the same interrogation time. The atomic resonance frequency is then determined with a higher resolution. For atomic clock applications, Ramsey technique has been first introduced in Cs beam tubes \cite{ramsey1956}, then in cold-atom fountains (see \cite{wynands_2005} and reference therein)  and more recently also in vapor cell arrangements \cite{godone2006, gozzelino2023, almat2020, shen2020}.

\begin{figure}[h!]
\includegraphics[width=0.9\columnwidth]{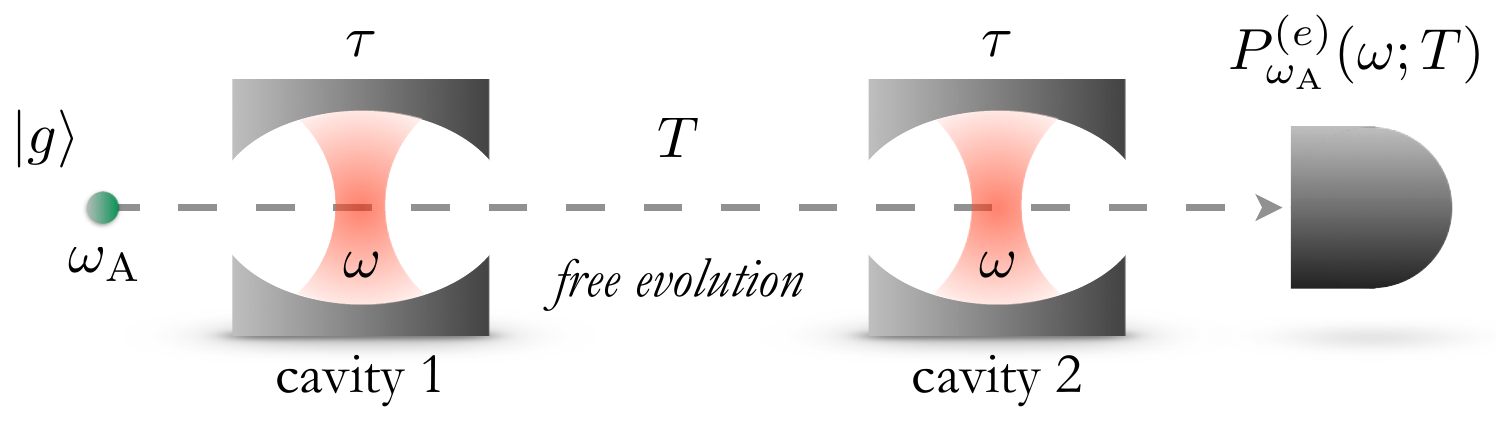}
\caption{Scheme of the Ramsey method.}\label{f:ramsey}
\end{figure}

The scheme of the Ramsey method is reported in Fig.~\ref{f:ramsey}: with respect to the Rabi scheme addressed in the previous section, now,
after the interaction with a first cavity, the atom undergoes a free evolution for a time $T$
and, then, it interacts with a second cavity before the final measurement. The interaction time is set to $\tau = \pi/(2\Omega_0)$ for both the cavities (note that if $T \to 0$ we recover the Rabi scheme) and, after the whole evolution, the probability to find the atom in the excited state reads (see Appendix~\ref{app:Ramsey} for the derivation):
\begin{align}
\Pe&(\omega;T) = \frac{4}{\left[ \Theta_{\Omega_0}(\Delta \omega)\right]^2}
\sin^2 \left[
\frac{\pi}{4} \Theta_{\Omega_0}(\Delta \omega)
\right] \nonumber\\[1ex]
&\times\Bigg\{
\cos\left(
\frac{   \kappa \, \Delta\omega }{4\, \Omega_0}
\right)
\cos\left[
\frac{\pi}{4} \Theta_{\Omega_0}(\Delta \omega)
\right] \nonumber\\[1ex]
& - \frac{\left(\Delta\omega  /  \Omega_0\right)}{\Theta_{\Omega_0}(\Delta \omega)}
\, \sin\left(
\frac{   \kappa \, \Delta\omega }{4\, \Omega_0}
\right)
\sin\left[
\frac{\pi}{4} \Theta_{\Omega_0}(\Delta \omega)
\right]
\Bigg\}^2\,,\label{eq:ramsey:pe}
\end{align}
where $\Theta_{\Omega_0}(\Delta \omega)$ is still given by Eq.~(\ref{Theta}).

\begin{figure}[tbh!]
\includegraphics[width=0.7\columnwidth]{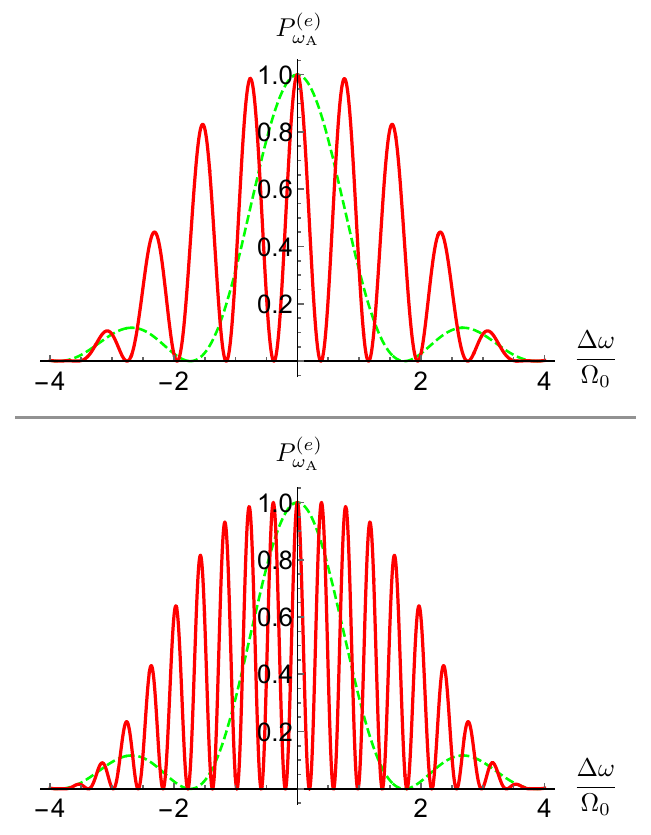}
\vspace{-0.2cm}
\caption{The probability $\Pe$ to find the atom in the excited state after the whole
evolution as a function of $\Delta\omega/\Omega_0$ (solid, red lines).
The atom is initially in the ground state and the interaction time $\tau$ with the cavity
field corresponds to a $\pi/2$-pulse. In the top and in the bottom plots we set the free
evolution time $T=5\, \tau$ and $T=10\, \tau$, respectively. The green dashed lines
refer to $\pe$ given in Fig.~\ref{f:rabi:pe}.}\label{f:ramsey:Pe}
\end{figure}

In Fig.~\ref{f:ramsey:Pe} we plot $\Pe$ for two different values of $T$: the
typical Ramsey interference fringes are evident and the longer the free evolution
time $T$ between the cavities, the larger the number of peaks. For comparison
in the same figure we reported $\pe$ (green dashed lines).

\begin{figure}[tb]
\includegraphics[width=0.7\columnwidth]{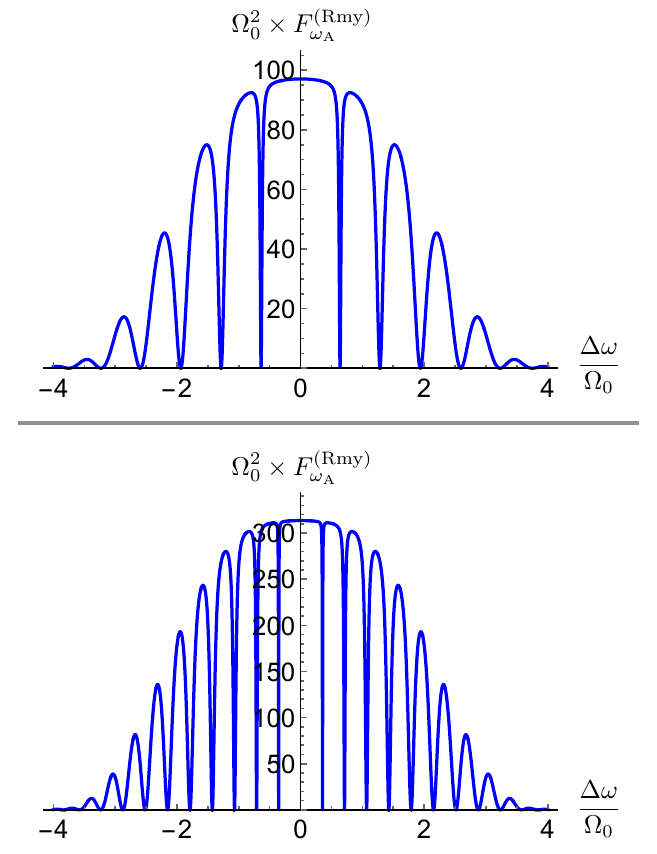}
\vspace{-0.2cm}
\caption{Plot of the Fisher information $\Fry$ associated with the Ramsey method as a
function of $\Delta\omega/\Omega_0$. The maximum is achieved at
$\omega = \omegaA$. In the top and in the bottom plots we set the free
evolution time $T=5\, \tau$ and $T=10\, \tau$, respectively. }\label{f:ramsey:FI}
\end{figure}

Since the width of the peak of $\Pe$ at $\omega = \omega_{\rm A}$ is smaller
than the one of $\pe$, we also expect the Fisher information
\begin{align}
\Fry (\omega) &= \sum_{k=g,e} \Pk(\omega;T) \left[ \partial_{\omegaA} \ln \Pk(\omega;T)\right]^2\\
&=  \frac{\left[ \partial_{\omegaA} \Pe(\omega)\right]^2}{\Pe(\omega) \left[1-\Pe(\omega) \right]}\,
\label{F:ramsey}
\end{align}
to be larger. Now, the analytic expression of $\Fry (\omega)$ is clumsy
and it is not reported here explicitly, but it is plotted in Fig.~\ref{f:ramsey:FI}
for the same values of $T$ chosen in Fig.~\ref{f:ramsey:Pe}.

We can see that the maximum of the Fisher information is still attained at $\omega = \omega_{\rm A}$.
If $\omega \approx \omega_{\rm A}$, we find the following expansion:
\begin{align}
\Fry &(\omega) \approx \frac{1}{\Omega_{0}^2} \Bigg\{
4\left(1 + \frac{\pi}{4}\, \kappa\right)^2 \nonumber\\[1ex]
&- \left[ 8- \frac{3}{4}\, \pi^2
+ (10 - 3 \pi) \frac{\pi}{2}\, \kappa \right] ( \omega - \omega_{\rm A} )^2
\Bigg\}\,,\label{Fry:expansion}
\end{align}
where, for the sake of simplicity, we set $T = \kappa \tau$ with $\tau = \pi/(2 \Omega_0)$.
Note that for $\kappa = 0$,  $\Fry (\omega)$ reduces to $\Fri (\omega)$ in Eq.~(\ref{Fri:expansion}).
Remarkably, if the free evolution time $T = \kappa \tau$ is much longer than the in-cavity interaction time $\tau$,
namely, $\kappa \gg 1$, Eq.~(\ref{Fry:expansion}) reduces to (still for $\omega \approx \omega_{\rm A}$):
\begin{align}
\Fry &(\omega) \approx \frac{4}{\Omega_{0}^2}
\left(1 + \frac{\pi}{4}\, \kappa\right)^2\,,
\end{align}
that is independent of $\omega$. It is also worth noting that, in the correspondence of the maximum
($\omega = \omega_{\rm A}$), the ratio between the Fisher information in the case of the Ramsey
and Rabi method, respectively, scales as:
\begin{equation}
\frac{\Fry (\omega_{\rm A})}{\Fri (\omega_{\rm A})} =
\left(1 + \frac{\pi}{4}\, \kappa\right)^2\,.
\end{equation}

\section{Quantum Fisher information for Rabi and Ramsey methods}\label{s:QFI}
\begin{figure}[tb]
\includegraphics[width=0.7\columnwidth]{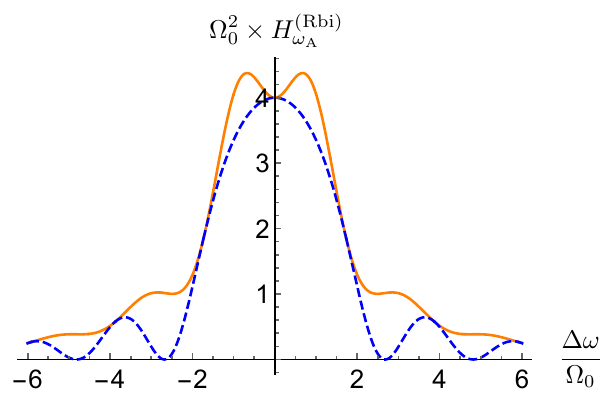}
\vspace{-0.3cm}
\caption{Plot of the quantum Fisher information $\QF^{\rm (Rbi)}_{\omegaA}$ associated with the Rabi method (orange solid line)
as a function of $\Delta\omega/\Omega_0$. The blue dashed line is the Fisher information
of the right panel of Fig.~\ref{f:rabi:pe}. Note that for $\omega = \omegaA$, the Fisher information and the quantum
Fisher information have the same value.}\label{f:rabi:QI}
\end{figure}

\begin{figure}[tb]
\includegraphics[width=0.7\columnwidth]{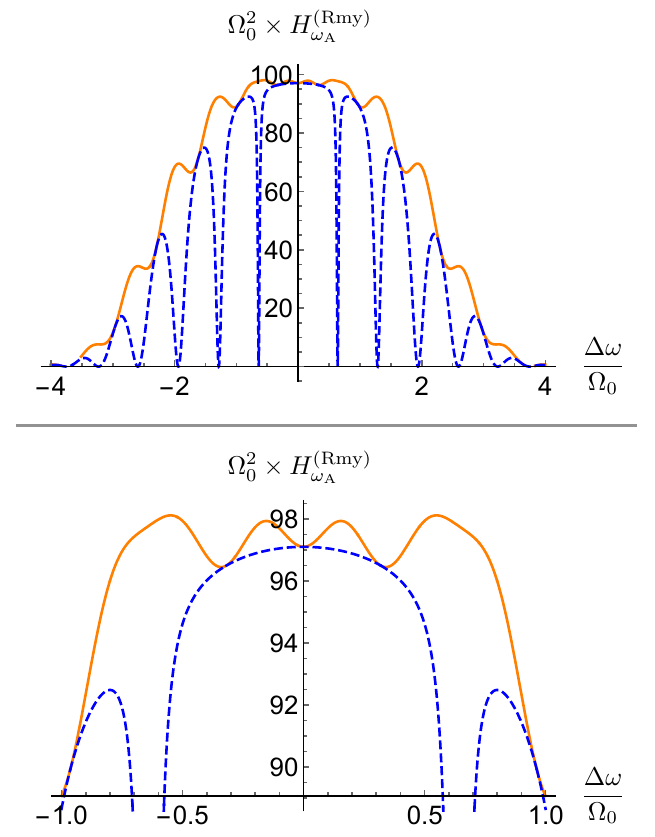}
\vspace{-0.2cm}
\caption{(Top) Plot of the quantum Fisher information $\QF^{\rm (Rmy)}_{\omegaA}$ associated with the
Ramsey method (orange solid line) as a function of $\Delta\omega/\Omega_0$. The blue dashed line is the Fisher information
of Fig.~\ref{f:ramsey:FI}. We set the free evolution time $T=5\, \tau$. (Bottom) Magnification
of the top plot. Note that for $\omega = \omegaA$, the Fisher information and the quantum
Fisher information have the same value.}\label{f:ramsey:QI}
\end{figure}

By explicitly evaluating Eq.~(\ref{QFI:def}) in the presence of the corresponding evolved states (see Appendices~\ref{app:Rabi} and \ref{app:Ramsey} for their explicit calculation) we retrieve
the quantum Fisher information for both the measurements, which are reported in Figs.~\ref{f:rabi:QI} and
\ref{f:ramsey:QI}, respectively. We note that for $\Delta \omega = 0$ the values of the Fisher and of the
quantum Fisher information are the same: in this regime the atomic population measurement turns out
to be optimal. However, we can also see that for $\Delta\omega \ne 0$ the quantum Fisher information
may be greater than the classical one obtained from the atomic population detection.

\section{The coherent population trapping (CPT) approach }\label{s:CPT}

\begin{figure}
\includegraphics[width=0.35\columnwidth]{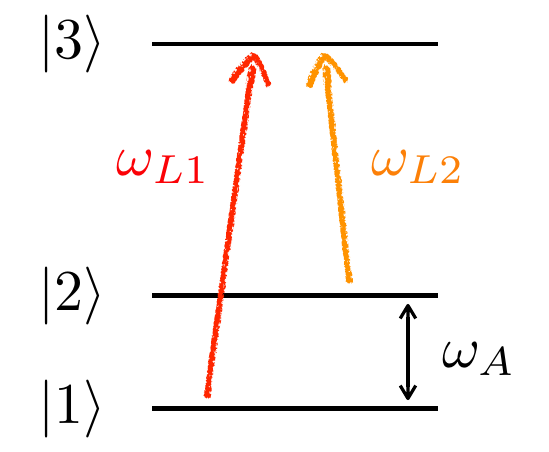}
\vspace{-0.3cm}
\caption{Three-level system interacting with a couple of phase-coherent laser fields in a $\Lambda$ scheme.}\label{three_level}
\end{figure}

The Coherent Population Trapping (CPT) is a quantum phenomenon in which two metastable levels (that could be the two levels defining a clock transition) are coupled to a common excited state through two quasi-resonant phase-coherent laser fields \cite{arimondo_1996}, as shown in Fig.~\ref{three_level}.
The trapping is made evident when the laser frequency difference $\omega_{L1}-\omega_{L2}$ matches the atomic frequency $\omegaA$, the so called Raman resonance condition. In this case, the atomic population is trapped in a coherent superposition of the two clock levels; the atoms are no longer able to absorb photons from the laser fields and consequently the fluorescence signal appears strongly reduced (dark line). For frequency standards applications, CPT has been implemented in cold atom setups but above all in vapor cell clocks using hot atoms.

With a density matrix approach, it is possible to evaluate the atomic populations and the coherence generated by the CPT phenomenon in the ground state. The free Hamiltonian of the system can be written in the matrix form as:
\begin{equation}\label{CPT:Ham}
{\cal H}_{\CPT} = \hbar\, \left(
\begin{array}{ccc}
 \omega_1 & 0 & \Omega_{\rm R1} \\[1ex]
0 & \omega_2 & \Omega_{\rm R2} \\ [1ex]
\Omega_{\rm R1} & \Omega_{\rm R2} & \omega_3
\end{array}
\right),
\end{equation}
where $\omega_k$ is the frequency associated with the $k$-th level, $k=1,2,3$, while $\Omega_{\rm R1}$ and $\Omega_{\rm R2}$ are the Rabi frequencies connected to the coupling between the levels $|1 \rangle$ and $|2 \rangle$ and the level $|3 \rangle$. Here for simplicity, we consider a symmetric $\Lambda$ scheme, in which the same Rabi frequency $\Omega_0 = \Omega_{\rm R1} = \Omega_{\rm R2}$.

Assuming the rotating wave and the adiabatic approximations and the absence of saturation, in the steady state conditions \cite{levi_2000}
the density operator $\hvarrho$ of the systems has the following matrix elements $\rho_{j,k} = \langle j | \hvarrho | k \rangle$, $j,k = 1,2,3$:
\begin{subequations}\label{CPT:eqs}
\begin{align}
\rho_{1,1} &= \rho_{2,2} = \frac{1}{2} (1 - \rho_{3,3}), \\[1ex]
\rho_{3,3}&= \frac{2\Gamma_p}{\Gamma^*}\left[1-\frac{2\Gamma_p(\gamma_2+2\Gamma_p)}{(\gamma_2+2\Gamma_p)^2+(\Delta\omega)^2}\right],\label{rho:33}\\[1ex]
\rho_{1,2}&=-\frac{\Gamma_p}{\gamma_2+2\Gamma_p+i \Delta\omega},\\[1ex]
\rho_{1,3}&=\rho_{2,3}=0.
\end{align}
\end{subequations}
We recall that $\rho_{k,k}$ represents the population of the $k$-th level whereas $\rho_{j,k}$, with $j\ne k$, corresponds to the coherences between the levels $j$ and $k$.
Here $\Delta\omega$ is the two-photon Raman detuning defined as:
\begin{equation}\label{Delta:omega:CPT}
    \Delta\omega= (\omega_{L1} - \omega_{L2}) - \omegaA
\end{equation}
where, $\omega_{L1}$ and  $\omega_{L2}$ are the (angluar) frequencies of the laser coupling levels $|1\rangle$ and $|3\rangle$ and levels $|2\rangle$ and $|3\rangle$, respectively, see Fig.~\ref{three_level}. Since the CPT approach is widely adopted in frequency standards operating with hot vapor cells, in previous equations we introduced the relaxation rates of the ground state coherence and of the excited state, $\gamma_2$ and $\Gamma^*$, respectively, which take into account the decoherence processes occurring in the cell \cite{vanier_audoin}. The laser pumping rate is given by $\Gamma_p=\Omega_0^2/(2\Gamma^*)$.

\begin{figure}[t!]
\includegraphics[width=0.35\columnwidth]{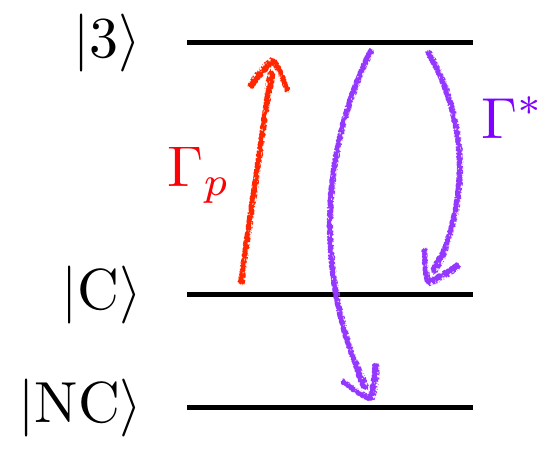}
\vspace{-0.3cm}
\caption{Lambda scheme in terms of the coupled, $|\C\rangle$, and uncoupled, $|\NC\rangle$, states; this represents an optical pumping situation in which the atoms are pumped in the dark state.}\label{three_level_nc}
\end{figure}
 The CPT physics can be better captured by describing the system using the so-called coupled, $|\C\rangle$, and uncoupled, $|\NC\rangle$, states \cite{scully}. In the case of symmetric $\Lambda$ scheme, they reads:
\begin{subequations}
\begin{align}\label{cpt_solutions}
|\C\rangle &=  \frac{1}{\sqrt{2}}(|1\rangle + |2\rangle), \\[1ex]
|\NC\rangle &=  \frac{1}{\sqrt{2}}(|2\rangle -|1\rangle ). 
\end{align}
\end{subequations}
It is possible to prove that the state $|\NC\rangle$ is not coupled to the radiation field \cite{scully}, thereafter, the following matrix element is identically null, namely:
\begin{equation}
    \langle3|{\cal H}_{\CPT}|\NC\rangle=0,
\end{equation}
whereas
\begin{equation}
    \langle3|{\cal H}_{\CPT}|\C\rangle=\frac{\hbar \Omega_{R}}{\sqrt{2}}.
\end{equation}
Therefore, if the atom is in the $|\NC\rangle$ state it cannot be excited to level $|3\rangle$ and, consequently, no fluorescence signal can be generated: whatever is the initial condition, the CPT Hamiltonian ${\cal H}_{\CPT}$ pumps all the atoms in the $|\NC\rangle$ state. The situation is represented in Fig.~\ref{three_level} and in Fig.~\ref{three_level_nc} in terms of the levels basis $\{ |1\rangle, |2\rangle, |3\rangle \}$ and coupled and non-coupled basis $\{ |\C\rangle, |\NC\rangle, |3\rangle \}$, respectively. We are in the presence of an optical pumping process, where the $|\NC\rangle$ state is populated by the relaxation from the excited state $|3\rangle$ and from the depopulation pumping from level $|\C\rangle$, which is coupled to the radiation field \cite{scully}. 

We can use Eqs.~(\ref{cpt_solutions}) to calculate the matrix elements expressed in the basis $|\NC\rangle$, $|\C\rangle$ and $|3\rangle$. In particular, we are interested in $\rho_{\NC,\NC}$ which gives the atomic population of the non-coupled state. It is straightforward to see that (for the sake of compactness we put $\rho_{l,l} = \rho_l$):
\begin{align}
    \rho_{\C}&=\frac{1}{2}\big(1-\rho_{3}+2\, \Re\mathrm{e}[\rho_{1,2}]\big),\\[1ex]
    \rho_{\NC}&=\frac{1}{2}\big(1+\rho_{3}-2\, \Re\mathrm{e}[\rho_{1,2}]\big),
\end{align}
which depend on the coherences $\rho_{1,2}$ between the levels $|1\rangle$ and $|2\rangle$,
and Eqs.~(\ref{CPT:eqs}) then yield:
\begin{align}
    \rho_{\C}=\frac{1}{2}\Bigg\{1 &+ \frac{2\Gamma_p}{\Gamma^*}\Bigg[1+\frac{2\Gamma_p(\gamma_2+2\Gamma_p)(\Gamma^*+2\Gamma_p)}{(\gamma_2+2\Gamma_p)^2+(\Delta\omega)^2}\Bigg]\Bigg\}\,.
   \label{rho:C:C}
\end{align}
and
\begin{align}
    \rho_{\NC}=\frac{1}{2}\Bigg\{1 &+ \frac{2\Gamma_p}{\Gamma^*}\Bigg[1+\frac{2\Gamma_p(\gamma_2+2\Gamma_p)(\Gamma^*-2\Gamma_p)}{(\gamma_2+2\Gamma_p)^2+(\Delta\omega)^2}\Bigg]\Bigg\}\,.
   \label{rho:NC:NC}
\end{align}

\begin{figure}[tb]
\includegraphics[width=0.7\columnwidth]{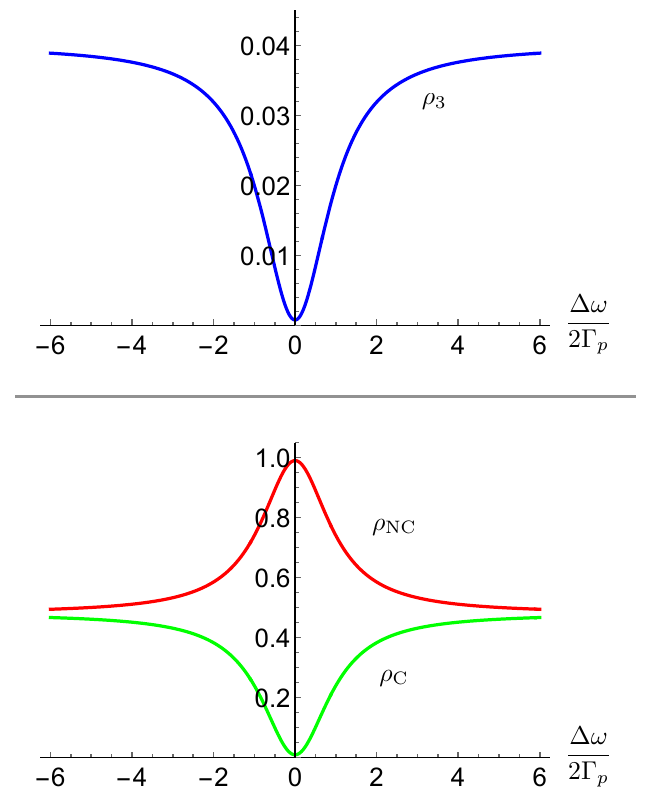}
\vspace{-0.2cm}
\caption{Plots of $\varrho_{3}$ (top) and of $\rho_{\C}$ and $\rho_{\NC}$ (bottom) as functions of the rescaled Raman detuning $\Delta\omega/ (2\Gamma_p)$ (see the text for details). We consider the typical values $\gamma_2 = 400~{\rm s}^{-1}$ and $\Gamma_p = 10^4~{\rm s}^{-1}$.}\label{fig:rho_NC_NC}
\end{figure}
The behavior of $\rho_{3}$, $\rho_{\C}$ and $\rho_{\NC}$ versus the rescaled Raman detuning $\Delta\omega/ (2\Gamma_p)$ is shown in the top and bottom panels of Fig.~\ref{fig:rho_NC_NC}, respectively. As expected, at resonance ($\Delta\omega=0$) almost all the atomic population is pumped into the uncoupled state $|\NC\rangle$, that is $\rho_{\NC} \approx 1$ and $\rho_{\C}, \rho_{3} \approx 0$.

First of all, we evaluate the Fisher information associated with the measurement of the population $\rho_l(\Delta\omega)$ of the level $| l \rangle$, with $l=\C, \NC, 3$, where we put in evidence the dependence on $\Delta\omega$ given in Eq.~(\ref{Delta:omega:CPT}):
\begin{align}
\FCPT (\Delta\omega) =  \sum_{l=\C, \NC, 3} \rho_l (\Delta \omega) \big[ \partial_{\omegaA} \ln \rho_l (\Delta\omega)\big]^2 \,.
\label{F:CPT}
\end{align}
Analogously, we can evaluate the Fisher information associated with the measurement involving the atomic energy states $| n \rangle$, with $n=1, 2, 3$:
\begin{align}
\FCPTen (\Delta\omega) =  \sum_{l=1,2,3} \rho_l (\Delta \omega) \big[ \partial_{\omegaA} \ln \rho_l (\Delta\omega)\big]^2 \,.
\label{F:CPT:123}
\end{align}
Finally, thanks to Eq.~(\ref{QFI:def}) and by using Eqs.~(\ref{CPT:eqs}) we obtain the corresponding quantum Fisher information $\HCPT(\Delta\omega)$, whose analytic expression is rather cumbersome and it is not reported explicitly (see Appendix~\ref{app:QFI:CPT} for further details). We recall that, by definition, the quantum Fisher information does not depend on the particular basis chosen to describe the system.

\begin{figure}[tb]
\includegraphics[width=0.7\columnwidth]{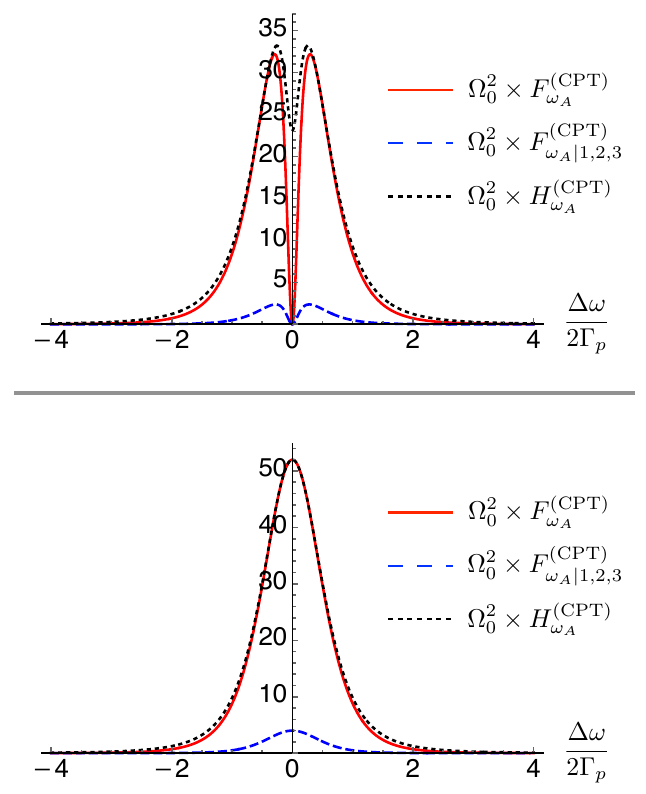}
\vspace{-0.2cm}
\caption{(Top) Plot of the Fisher information $\FCPT$ (red, solid line) and $\FCPTen$ (blue, dashed line)  as functions of $\Delta\omega / (2\Gamma_p)$. We set $\gamma_2 = 400~{\rm s}^{-1}$ and $\Gamma_p = 10^4~{\rm s}^{-1}$. (Bottom) Same plots of the top panel but now we set $\gamma_2 \to 0$, that is we neglect the relaxation of the ground states coherence: the maximum is achieved at $\omega_{L1} - \omega_{L2} = \omegaA$. See the text for details.}\label{f:CPT:Fisher}
\end{figure}
In Fig.~\ref{f:CPT:Fisher} we plot $\FCPTlev$ as a function of $\Delta\omega / (2\Gamma_p)$: we can see that the maximum value of $\FCPTlev$ is similar that obtained with the Rabi method (see Fig.~\ref{f:rabi:pe}). It is clear that the information gained by the measurement of the energy levels (blue dashed lines in the figure) is much less with respect to that retrieved by measuring the system in the coupled/uncoupled basis. Therefore, performing a measurement that involves coherences between the energy states $|1\rangle$ and $|2\rangle$, as the assessment of the density matrix elements $\rho_\C$ and $\rho_\NC$, allows retrieving a higher Fisher information that, in the limit of very small relaxation rate $\gamma_2$ and at resonance ($\Delta \omega = 0$), reaches the maximum value given the quantum Fisher information, as highlighted in the bottom panel of Fig.~\ref{f:CPT:Fisher}.
Concerning the practical point of view, we point out that the maser approach represents a technique to detect the clock transition through the atomic coherence \cite{godone2000, godone2006b}.

Remarkably, in the top panel of the Fig.~\ref{f:CPT:Fisher} we can clearly see the presence of a deep at $\Delta \omega = 0$: this is an ``accident'' that follows from the definition of the Fisher information. In fact, while the derivative with respect to $\omega_A$ of all the elements $\rho_l$ vanishes at resonance (there is a maximum or a minimum), as one can see, for instance, in Fig.~\ref{fig:rho_NC_NC}, in realistic conditions none of these reaches 1 or 0 due to the presence of the relaxation of the ground states coherence $\gamma_2 > 0$. Hence, the Fisher information vanishes and the quantum Fisher information displays a deep as well. Nevertheless, in actual experiments one can post-process the data to ``normalize'' the measured quantities: the effect of this ``normalization'' is that the elements $\rho_l$ now can reach 1 or 0 and the Fisher information no longer goes to zero at resonance but reaches a maximum, as expected. This can be seen in the bottom panel of Fig.~\ref{f:CPT:Fisher} where we neglected the relaxation of the ground states coherence by setting $\gamma_2 = 0$, that is mathematically equivalent to suitably ``normalize'' the matrix elements $\rho_l$, as one can see in particular form Eq.~(\ref{rho:33}).

\begin{figure}[t!]
\includegraphics[width=0.7\columnwidth]{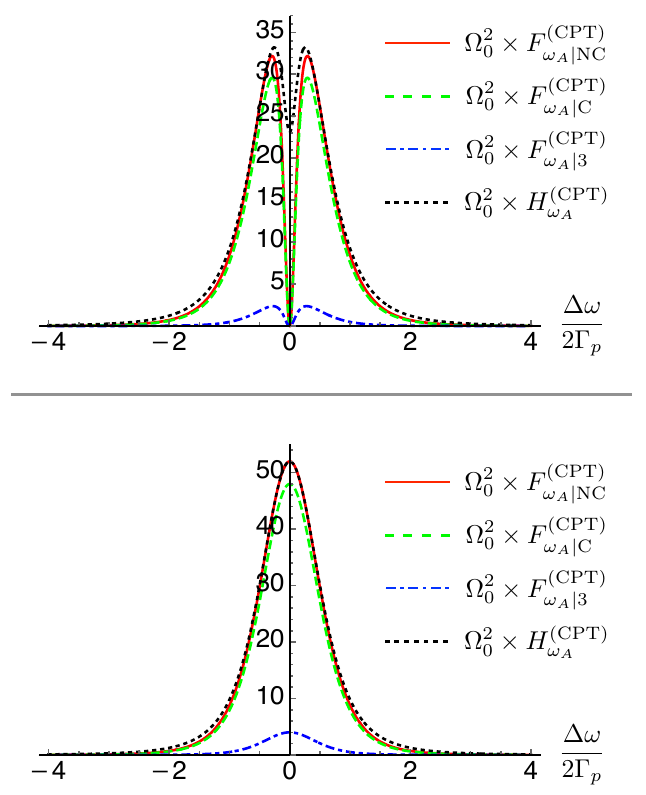}
\vspace{-0.2cm}
\caption{(Top) Plot of the single-level Fisher information $\FCPTlev$, $l=\C, \NC, 3$,  as functions of $\Delta\omega / (2\Gamma_p)$. We set $\gamma_2 = 400~{\rm s}^{-1}$ and $\Gamma_p = 10^4~{\rm s}^{-1}$. (Bottom) Same plots of the top panel but now we set $\gamma_2 = 0$, that is we neglect the relaxation of the ground states coherence:
as expected, the maximum is achieved at
$\omega_{L1} - \omega_{L2} = \omegaA$. See the text for details. }\label{f:CPT:Fisher:single}
\end{figure}
To better investigate the latter considerations, we can evaluate the Fisher and the quantum Fisher information associated with the measurement of just one of the states $| l \rangle$, with $l=\C, \NC, 3$ (analogous results are obtained addressing the energy-level states). Since, now, the measurement outcome is dichotomic, the Fisher information can be written as:
\begin{align}
\FCPTlev (\Delta\omega) =  \frac{\big[ \partial_{\omegaA} \rho_{l}  (\Delta\omega) \big]^2}{\rho_{l} (\Delta\omega) \big[1-\rho_{l} (\Delta\omega) \big]}\,.
\label{F:CPT:X}
\end{align}
This makes evident that, at resonance, $\FCPTlev$ vanishes since $\partial_{\omegaA} \rho_{l}  (0) = 0$ but $\rho_{l} (0) \left[1-\rho_{l} (0) \right] \ne 0$. If we set $\gamma_2 = 0$ or, equivalently, suitably normalize $\rho_{l}$ as discussed above, we still have $\partial_{\omegaA} \rho_{l}  (0) = 0$ while, now, $\rho_{l} (0) \left[1-\rho_{l} (0) \right] = 0$ and the Fisher information reaches a maximum as shown in Fig.~\ref{f:CPT:Fisher:single}.

\section{Conclusions}\label{s:conclusion}
In this paper, we have investigated the ultimate quantum limits to the achievable uncertainty in the estimation of the transition frequency between two atomic levels.

We have used classical and quantum estimation theory to assess the three most important techniques in this field, and proved that that measuring the atomic population allows one to reach the minimum uncertainty. We 
have also shown that the Ramsey method beats both 
the Rabi and the CPT performance, confirming the current  experimental evidence and justifying it within the framework of estimation and quantum estimation theory. Moreover, we have found that for CPT the Fisher and the quantum Fisher information coincide for any value of the detuning $\Delta \omega = \omegaA - \omega$, whereas for the Rabi and Ramsey methods, the quantum Fisher information is sensibly higher that the classical one only for nonzero values of $\Delta \omega$. We believe that this can foster new investigation in the search of different detection schemes able to reach that limit addressing also the coherences between the involved levels.

Even if the Ramsey method results more effective in terms of Fisher information compared to Rabi and CPT techniques, it is nevertheless evident that some applications require very small devices (see for example \cite{kitching}) for which CPT is particularly suited, since the clock transition is excited without the need of any microwave cavity. Also, in optical clocks the Rabi interaction is widely used, being much more easier to implement than the Ramsey one. In this paper, however, we analyzed the three techniques from a mere statistical point of view, regardless for the implementation and practical uses of the clock.

\acknowledgments
The authors acknowledge useful discussions with Marco Mancini in the early stage of this work. This work has been partially supported by the Italian Ministry of Research and Next Generation EU via the 
PRIN-2022 project RISQUE (contract n. 2022T25TR3) and the NQSTI-Spoke2-BaC project QMORE (contract n. PE00000023-QMORE).

\begin{appendix}

\section{Derivation of Eq.~(\ref{eq:Rabi:pe})}\label{app:Rabi}

A two-level atom, with transition frequency $\omegaA$ between the ground ($ | g \rangle$)
and the excited ($ | e \rangle$) states, is described by the Hamiltonian
\begin{equation}
\HA = \frac{\hbar \omegaA}{2} \, \sigmaZ ,
\end{equation}
$\sigmaZ = | e \rangle \langle e | - | g \rangle \langle g |$ being the Pauli matrix.
We assume that the atom interacts with an oscillating electric field
$\vec{E}(t) = E_0 \, \varepsilonF \, (e^{-i\omega t - i \phi} + e^{i\omega t + i \phi})$
inside a cavity, where $E_0$, $\omega$ and $\varepsilonF$ are the amplitude, the frequency and the polarisation
of the field. If we assume the dipole approximation, the interaction between atom and field can be
described by the following Hamiltonian:
\begin{align}
\Hint &= - \hat{D}\cdot \vec{E}(t)
\end{align}
where $\hat{D} = \vec{d}\, ( \sigma_{+} + \sigma_{-} )$,
$\vec{d}$ being the dipole transition moment, whereas $\sigma_{+} = | e \rangle \langle g |$ and
$\sigma_{-} = | g \rangle \langle e |$ are the rising and lowering operator, respectively.
\par
Upon introducing the detuning $\Delta \omega = \omegaA - \omega$, the total Hamiltonian can be
written as:
\begin{align}\label{rabi}
H = \frac{\hbar \Delta\omega}{2} \, &\sigmaZ + \frac{\hbar \omega}{2} \, \sigmaZ\nonumber \\[1ex]
&- \frac{\hbar \Omega_0}{2} \, ( \sigma_{+} + \sigma_{-} )(e^{-i\omega t - i \phi} + e^{i\omega t + i \phi}),
\end{align}
where $\Omega_0 = 2 \vec{d}\cdot \varepsilonF E_0 / \hbar$ is the Rabi frequency.
\par
In the interaction picture with respect to the Hamiltonian:
\begin{align}
H_0 = \frac{\hbar \omega}{2} \, \sigmaZ,
\end{align}
and performing the rotating wave approximation, i.e. neglecting the terms proportional to
$e^{\pm 2i \omega t}$, Eq.~(\ref{rabi}) becomes:
\begin{subequations}
\begin{align}\label{rabi1}
\tilde{H} &= \frac{\hbar \Delta\omega}{2} \, \sigmaZ - \frac{\hbar \Omega_0}{2} \, ( \sigma_{+} e^{- i \phi}  + \sigma_{-} e^{i \phi}),\\[1ex]
&= \frac{\hbar \Omega}{2} \, \vec{n}\cdot \vec{\sigma},
\end{align}
\end{subequations}
where $\vec{\sigma} = (\sigmaX, \sigmaY, \sigmaZ)$ is the vector of the Pauli matrices,
$\Omega = \sqrt{\Omega_0^2 + (\Delta \omega)^2}$ and:
\begin{equation}
\vec{n} = - \frac{1}{\Omega} \left(
\Omega_0 \cos\phi , \Omega_0 \sin\phi, - \Delta \omega
\right).
\end{equation}
Thereafter, the evolution operator (in the interaction picture) reads:
\begin{subequations}\label{cavity:evol}
\begin{align}
\tilde{U}(t) &= \exp\left( - i \frac{\tilde{H}}{\hbar} \, t \right)\,\\[1ex]
&= \left(\begin{array}{cc}
A_t(\Omega_0, \Delta\omega) & e^{-i\phi} B_t(\Omega_0, \Delta\omega) \\[1ex]
e^{i\phi} B_t(\Omega_0, \Delta\omega) & A_t^* (\Omega_0, \Delta\omega)
\end{array}
\right) \label{rabi:evol:matrix}
\end{align}
\end{subequations}
with:
\begin{subequations}
\begin{align}
A_t(\Omega_0, \Delta\omega) &= \cos\left( \frac{\Omega t}{2} \right)
- i \frac{\Delta\omega}{\Omega} \sin\left( \frac{\Omega t}{2} \right),\\[1ex]
B_t(\Omega_0, \Delta\omega) &= i \frac{\Omega_0}{\Omega} \sin\left( \frac{\Omega t}{2} \right).
\end{align}
\end{subequations}
In Eq.~(\ref{rabi:evol:matrix}) we used the matrix formalism, where
we considered as basis:
\begin{equation}
| g \rangle =  \left(\begin{array}{c}
0 \\ 1
\end{array}
\right)
\quad \mbox{and} \quad
| e \rangle =  \left(\begin{array}{c}
1 \\ 0
\end{array}
\right).
\end{equation}
Now, given the evolved state:
\begin{equation}\label{Uevol:Rabi}
| \psi(t) \rangle = \tilde{U}(t)| g \rangle,
\end{equation}
if we choose the interaction time such that
$\Omega_0 t = \pi$, namely, in the presence of a $\pi$-pulse, the probability
\begin{equation}
\pe(\omega) = | \langle e | \psi(t= \pi/\Omega_0) \rangle |^2
\end{equation}
is equal to Eq.~(\ref{eq:Rabi:pe}).

\section{Derivation of Eq.~(\ref{eq:ramsey:pe})}\label{app:Ramsey}

The evolution
through the cavities is still described by the evolution operator given in Eqs.~(\ref{cavity:evol}),
whereas the free evolution, in the interaction picture, is obtained applying the following
operator:
\begin{subequations}\label{free:evol}
\begin{align}
\tilde{U}_{\rm free}(t) &= \exp\left( - i \frac{\Delta\omega\, t}{2}\, \sigmaZ  \right)\,\\[1ex]
&= \left(\begin{array}{cc}
\exp\left( - i \frac{\Delta\omega\, t}{2}  \right) & 0 \\[1ex]
0 & \exp\left( i \frac{\Delta\omega\, t}{2}  \right)
\end{array}
\right)\,.
\end{align}
\end{subequations}
If the interaction time is $\tau$ for both the cavities, the whole evolution can be written as:
\begin{equation}\label{Uevol:Ramsey}
| \Psi_T(\tau) \rangle = \tilde{U}_{2}(\tau)\tilde{U}_{\rm free}(T)\tilde{U}_{1}(\tau)| g \rangle
\end{equation}
where $\tilde{U}_{k}(\tau)$ is given in Eqs.~(\ref{cavity:evol}) with $\phi = \phi_k$ and,
without loss of generality, we can assume $\phi_1 = 0$ and $\phi_2 = \phi$.
Now, if we now assume $\phi = 0$, $\tau = \pi/(2\Omega_0)$ (that is a $\pi/2$-pulse)
and $T = \kappa \tau$, the probability to find the atom in the state $| e \rangle$ after the whole evolution is:
\begin{equation}
\Pe(\omega; T) = | \langle e | \Psi_T(\tau=\pi/(2\Omega_0)) \rangle |^2\,,
\end{equation}
that leads to Eq.~(\ref{eq:ramsey:pe}).

\section{On the derivation of the quantum Fisher information from the CPT density matrix}\label{app:QFI:CPT}

To evaluate the quantum Fisher information from Eq.~(\ref{QFI:def}) it is enough finding the eigenvectors $| \psi_l \rangle$ and the eigenvalues $r_l$, $l=1,2,3$, of the density matrix $\hvarrho$ whose matrix elements are given by Eqs.~(\ref{CPT:eqs}). We have:
\begin{align}
| \psi_1 \rangle &= \frac{\Big(1 - i \DeltaSte\Big) | 1 \rangle +\sqrt{1 + \DeltaSte^2} | 2 \rangle }{\sqrt{2(1 + \tilde{\Delta}^2)}},  \\[1ex]
| \psi_2 \rangle &= \frac{\Big(1 + i \DeltaSte\Big) | 2 \rangle - \sqrt{1 + \DeltaSte^2} | 1 \rangle }{\sqrt{2}(1 + i \DeltaSte)}, \\[1ex]
| \psi_3 \rangle &= | 3 \rangle,
\end{align}
where $| l \rangle$, $l=1,2,3$, are the three energy levels of the atom and:
\begin{align}
\DeltaSte = \frac{\Delta\omega}{\gamma_2 + 2\Gamma_p}.
\end{align}
The corresponding eigenvalues, $\hvarrho | \psi_l \rangle = r_l | \psi_l \rangle$, are:
\begin{align}
r_1 &= \frac12 \left(
1 - \frac{2\GammaSte}{\sqrt{1+\DeltaSte^2}} - r_3
\right),  \\[1ex]
r_2 &= \frac12 \left(
1 + \frac{2\GammaSte}{\sqrt{1+\DeltaSte^2}} - r_3
\right), \\[1ex]
r_3 &= \frac{2\Gamma_p}{\Gamma^*}\,\frac{\gammaSte+\DeltaSte^2}{1 + \DeltaSte^2} ,
\end{align}
with
\begin{align}
\gammaSte = \frac{\gamma_2}{\gamma_2 + 2\Gamma_p},\quad
\GammaSte = \frac{\Gamma_p}{\gamma_2 + 2\Gamma_p}.
\end{align}

Finally, the analytic expression of the quantum Fisher information can be retrieved through straightforward calculations from:
\begin{equation}
\HCPT(\Delta\omega) = 2\, \sum_{m,n}
\frac{\big| \langle  \psi_n  |  \partial_{\omegaA} \hvarrho |  \psi_m  \rangle \big|^2}{r_n + r_m}\,,
\end{equation}
where $m,n = 1,2,3$. Due to its clumsy appearance, it is not reported here.  

\end{appendix}



\end{document}